\documentclass[a4paper,aps,prd,10pt,preprintnumbers,showpacs,twocolumn,superscriptaddress,nofootinbib,amsmath,amssymb]{revtex4-1}
\usepackage{graphicx}
\usepackage{cmap}
\usepackage[utf8]{inputenc}
\usepackage[T1]{fontenc}
\usepackage{hyperref}

\def\K{{\cal K}}

\begin{document}
\title{Two types of quasinormal modes of Casadio-Fabbri-Mazzacurati brane-world black holes}

\author{Bekir Can Lütfüoğlu}
\email{bekir.lutfuoglu@uhk.cz}
\affiliation{Department of Physics, Faculty of Science, University of Hradec Králové, Rokitanského 62/26, 500 03 Hradec Králové, Czech Republic}

\author{Sardor Murodov} 
\email{mursardor@ifar.uz}
\affiliation{New Uzbekistan University, Movarounnahr Str. 1, Tashkent 100000, Uzbekistan}
\affiliation{Institute of Fundamental and Applied Research, National Research University TIIAME, Kori Niyoziy 39, Tashkent 100000, Uzbekistan}

\author{Mardon Abdullaev} 
\email{mardonabdullaev@gmail.com}
\affiliation{Tashkent State Technical University, Tashkent 100095, Uzbekistan}
\affiliation{Kimyo International University in Tashkent, Shota Rustaveli street 156, Tashkent 100121, Uzbekistan}

\author{Javlon Rayimbaev} 
\email{javlon@astrin.uz}
\affiliation{Institute of Theoretical Physics, National University of Uzbekistan, Tashkent 100174, Uzbekistan}
\affiliation{University of Tashkent for Applied Sciences, Gavhar Str. 1, Tashkent 700127, Uzbekistan}

\author{Munisbek Akhmedov} \email{munisbek95@urdu.uz} \affiliation{Urgench State University, Kh. Alimjan Str. 14, Urgench 221100, Uzbekistan}

\author{Muhammad~Matyoqubov} 
 \email{m_matyoqubov@mamunedu.uz} 
\affiliation{Mamun University, Bolkhovuz Street 2, Khiva 220900, Uzbekistan}

\begin{abstract}
Using the convergent Leaver method, we investigate the quasinormal modes of a massive scalar field propagating in the background of the Casadio--Fabbri--Mazzacurati (CFM) brane-world black hole. We show that the spectrum exhibits two distinct types of modes, depending on their behavior as the field mass increases. In one class, the real oscillation frequency decreases and eventually approaches zero, while in the other the damping rate tends to vanish. When either the real or imaginary part of the frequency reaches zero, the corresponding mode disappears from the spectrum, and the first overtone replaces it. The emergence of modes with a vanishing real part at certain critical values of the field mass is a distinctive feature of the CFM spectrum.
\end{abstract}
\maketitle

\section{Introduction}

Quasinormal modes (QNMs) provide a fundamental tool for probing the geometry and physical properties of compact objects through their linear response to external perturbations \cite{Kokkotas:1999bd, Konoplya:2011qq, Berti:2009kk, Bolokhov:2025uxz}. In the context of black holes, the quasinormal spectrum governs the ringdown phase of gravitational-wave signals and is determined entirely by the background spacetime and the nature of the perturbing field \cite{LIGOScientific:2016aoc,LIGOScientific:2017vwq,LIGOScientific:2020zkf,Babak:2017tow}. In recent years, increasing attention has been paid to black holes in various alternative theories of gravity where the deviation from the Schwarzschild or Kerr solution in the ringdown phase could be observed.

Among such scenarios, a particularly well-motivated class is provided by brane-world models, where effective four-dimensional geometries emerge from higher-dimensional gravity  \cite{Randall:1999ee,Randall:1999vf,Dvali:2000hr}. 

Quasinormal spectra of black holes arising in higher-dimensional gravity theories and brane-world scenarios have been investigated from multiple perspectives over the past two decades \cite{Cardoso:2002pa,Morgan:2009pn,Konoplya:2007jv,Konoplya:2013sba,Cardoso:2003vt,Cho:2011sf,Emparan:2015rva,Han:2026fpn,Cuyubamba:2016cug,Arbelaez:2025gwj,Arbelaez:2026eaz}. Early and subsequent studies addressed wave dynamics and stability properties in a variety of effective geometries motivated by extra dimensions, warped spacetimes, and induced gravity on the brane \cite{Molina:2016tkr,Chen:2007jz,Zhidenko:2009zx,Yang:2014cra,BendasoliPavan:2006vk,Soleimani:2016mfh,daRocha:2017lqj,Abdalla:2006qj,Seahra:2005wk,Koyama:2005gh,Rahman:2022fay}. These works established that higher-dimensional effects can lead to nontrivial modifications of the effective potential and, consequently, of the QNM spectrum when compared to the four-dimensional Schwarzschild case.

Within this broad class of models, particular attention has been devoted to black holes localized on the brane, whose perturbative response encodes both four-dimensional and bulk-induced features. The quasinormal ringing of such brane-localised black holes has been analyzed in a number of studies employing both frequency- and time-domain techniques \cite{Kanti:2005xa,Chung:2015mna,Seahra:2005us,Zhidenko:2008fp,Zinhailo:2024jzt,Koyama:2005gh,Kanti:2006ua,Seahra:2005wk}. These analyses demonstrated that, despite significant similarities with standard four-dimensional black holes at early times, the presence of extra dimensions and bulk effects may leave observable imprints in the real oscillation frequency and damping rates of the evolution of perturbations.

In this framework, Casadio, Fabbri and Mazzacurati (CFM) proposed a family of exact solutions to the effective Shiromizu–Maeda–Sasaki equations, interpolating between black holes and traversable wormholes depending on a continuous parameter \cite{Casadio:2001jg}. This geometry has been extensively studied as a prototype of black-hole–wormhole transition spacetimes. The quasinormal ringing and time-domain response of these backgrounds were investigated in Refs.~\cite{Abdalla:2006qj,Abdalla:2007zz,Bronnikov:2019sbx}, where it was shown that, for massless test fields, the early-time signal closely mimics that of a black hole, while near the transition threshold the late-time dynamics may exhibit echoes characteristic of a wormhole geometry.

However, in all existing analyses of QNMs in this class of spacetimes, the perturbing fields were assumed to be {\it massless}. This restriction leaves open an important and physically well-motivated question: how does the presence of a mass term in the perturbation equation modify the quasinormal spectrum and the late-time behavior of the signal in brane-world black-hole–wormhole geometries?

Quasinormal modes of massive fields of various spin have been extensively studied in a wide range of gravitational backgrounds (see, for example, \cite{Konoplya:2004wg,Konoplya:2006br,Konoplya:2018qov,Konoplya:2017tvu,Zhidenko:2006rs,Konoplya:2005hr,Ohashi:2004wr,Zhang:2018jgj,Aragon:2020teq,Ponglertsakul:2020ufm,Gonzalez:2022upu,Burikham:2017gdm} and references therein). These studies have revealed several intriguing features that do not arise in the massless case. In particular, the presence of a mass term may lead to qualitatively new spectral phenomena when the field mass is tuned.

Firstly, effective mass terms naturally appear in perturbation equations in certain higher-dimensional scenarios due to the influence of the bulk on the brane \cite{Seahra:2004fg,Ishihara:2008re}. Secondly, massive gravitons, either in explicit massive gravity theories or as effective degrees of freedom, have been argued to contribute to very long-wavelength gravitational signals \cite{Konoplya:2023fmh}, which are currently being probed by Pulsar Timing Array experiments \cite{NANOGrav:2023gor,NANOGrav:2023hvm}. Thirdly, massive fields may support arbitrarily long-lived QNMs for particular values of the field mass, leading to quasi-resonant behavior \cite{Ohashi:2004wr,Konoplya:2004wg}. This phenomenon has been shown to be remarkably universal, occurring for different spins \cite{Konoplya:2005hr,Fernandes:2021qvr,Konoplya:2017tvu,Percival:2020skc}, black-hole backgrounds \cite{Konoplya:2006br,Konoplya:2013rxa,Bolokhov:2024bke,Lutfuoglu:2025hjy,Zhidenko:2006rs,Zinhailo:2018ska,Konoplya:2019hml,Bolokhov:2023bwm,Bolokhov:2023ruj,Lutfuoglu:2026fpx,Lutfuoglu:2025kqp,Lutfuoglu:2025qkt,Lutfuoglu:2025bsf,Skvortsova:2024eqi}, and even for other compact objects such as wormholes \cite{Churilova:2019qph,Lutfuoglu:2025hwh}. At the same time, the existence of arbitrarily long-lived modes is not guaranteed, and there are known examples where such modes do not occur despite the presence of a mass term \cite{Zinhailo:2024jzt,Konoplya:2005hr}.

Another important consequence of a nonzero field mass is the qualitative modification of late-time decay. When the quasinormal ringing is replaced by asymptotic tails, massive fields generically exhibit oscillatory late-time behavior instead of the standard power-law decay characteristic of massless perturbations. Such oscillatory tails have been extensively studied in various contexts \cite{Jing:2004zb,Koyama:2001qw,Moderski:2001tk,Konoplya:2006gq,Rogatko:2007zz,Koyama:2001ee,Koyama:2000hj,Gibbons:2008gg,Gibbons:2008rs}. Furthermore, even initially massless fields may effectively acquire a mass when propagating in certain external environments, such as in the vicinity of a black hole immersed in a magnetic field \cite{Konoplya:2007yy,Konoplya:2008hj,Wu:2015fwa,Siahaan:2025dbx,Davlataliev:2024mjl}.

Motivated by these considerations, in this paper, we extend previous analyses of the CFM brane-world geometry \cite{Abdalla:2006qj,Bronnikov:2019sbx,Malik:2024itg} by studying QNMs of a \emph{massive scalar field}. Previous studies of perturbations of the CFM black hole \cite{Bronnikov:2019sbx,Malik:2024itg} have been restricted to massless fields, while Ref.~\cite{Abdalla:2006qj} considered the massive case only in the context of asymptotic tails and primarily for a different model corresponding to zero-mass black holes. Thus, to the best of our knowledge, a systematic analysis of the quasinormal spectrum of massive field perturbations in the CFM background has not been performed so far. Our aim is to understand how the mass term modifies the quasinormal spectrum, damping rates, and late-time behavior of perturbations in a spacetime interpolating between black-hole and wormhole configurations. We show that the spectrum of massive field perturbations is qualitatively different not only from the massless case, but also from that of a massive field in the Schwarzschild spacetime. In particular, two distinct types of QNMs emerge: those whose damping rate vanishes at certain critical values of the mass parameter, and those whose real part vanishes at specific values of $\mu$. In the Schwarzschild/Kerr case only the former type is present, whereas for massless perturbations in the CFM background neither type appears.

The paper is organized as follows. In Sec.~II we briefly review the CFM brane-world black-hole solution and derive the wave equation governing massive scalar perturbations. In Sec.~III we outline the methods used for the computation of QNMs, including the WKB approximation and the convergent Leaver method. Section~IV is devoted to the analysis of the quasinormal spectrum and its dependence on the scalar-field mass and the tidal parameter. Finally, in Sec.~V we summarize our results and discuss their implications.

\section{The CFM black-hole spacetime and wave-like equation}\label{sec:wavelike}

Beginning with the five-dimensional vacuum Einstein equations and projecting them onto a four-dimensional hypersurface (the brane), one obtains an effective gravitational theory governing dynamics on the brane. Working in Gaussian normal coordinates $(x^{\nu}, z)$, where $\nu = 0,1,2,3$ label coordinates along the brane and $z$ parametrizes the extra dimension, the induced field equations on the brane assume the Shiromizu–Maeda–Sasaki (SMS) form \cite{Shiromizu:1999wj},
\begin{equation}\label{h-cons}
R^{(4)}_{\mu\nu} = \Lambda_{4}\, g^{(4)}_{\mu\nu} - E_{\mu\nu},
\end{equation}
where $R^{(4)}_{\mu\nu}$ and $g^{(4)}_{\mu\nu}$ denote the Ricci tensor and metric intrinsic to the brane, and $\Lambda_{4}$ represents the effective four-dimensional cosmological constant. The tensor $E_{\mu\nu}$ is the projection of the five-dimensional Weyl tensor onto the brane and is traceless by construction. It encapsulates nonlocal gravitational effects originating from the bulk and can be viewed as a tidal imprint of the higher-dimensional geometry on the brane spacetime.

Because $E_{\mu\nu}$ is not determined by four-dimensional quantities alone, the system is not fully closed at the level of Eq.~(\ref{h-cons}). However, certain combinations of these equations can be written in closed form. In particular, taking the trace eliminates $E_{\mu\nu}$ and yields \cite{Bronnikov:2003gx}
\begin{equation}\label{trace}
R^{(4)} = 4\,\Lambda_{4}.
\end{equation}
This relation plays the role of a Hamiltonian constraint in the Arnowitt–Deser–Misner (ADM) decomposition and serves as the primary condition restricting admissible four-dimensional geometries on the brane.

\begin{figure*}
\resizebox{\linewidth}{!}{\includegraphics{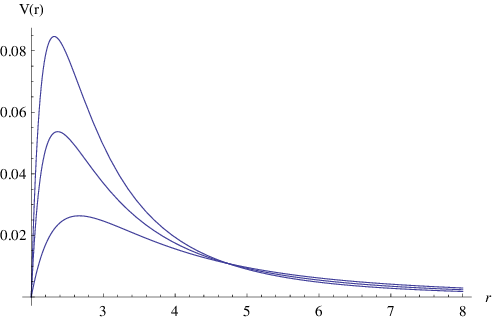}\includegraphics{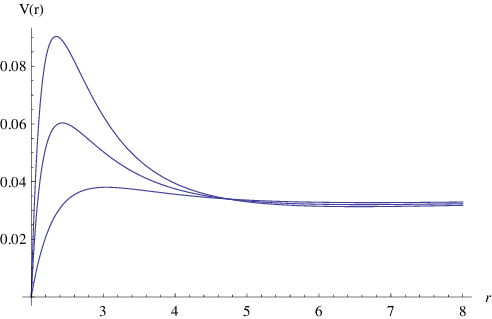}\includegraphics{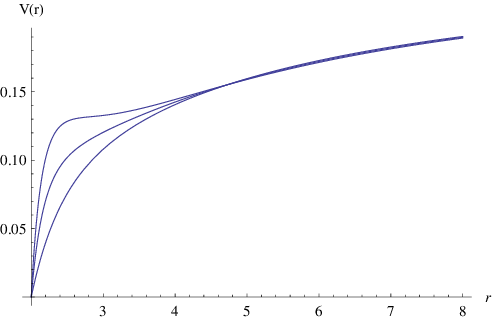}}
\caption{Left Panel: Effective potentials for $\ell=0$, $\mu=0$ perturbations ($M=1$): $\gamma=3$, $\gamma=0$, and $\gamma=-3$ from bottom to top. Middle Panel: The same for $\mu=0.2$. Right Panel: The same for $\mu=0.5$.}\label{fig:pot}
\end{figure*}

A static, spherically symmetric solution of Eq.~(\ref{trace}), known as the CFM spacetime, can be written as \cite{Casadio:2001jg}
\begin{equation}\label{metric}
ds^{2} = -f(r)\,dt^{2} + \frac{B^{2}(r)}{f(r)}\,dr^{2}
+ r^{2}\left(d\theta^{2} + \sin^{2}\theta\, d\phi^{2}\right),
\end{equation}
with
\[
f(r) = 1 - \dfrac{2M}{r}, 
\qquad
B^{2}(r) = \dfrac{1 - \dfrac{3M}{2r}}{1 - \dfrac{M\gamma}{2r}},
\]
where $M$ denotes the mass parameter and $\gamma$ is a dimensionless quantity characterizing the tidal influence of the bulk geometry. Throughout the remainder of the paper we express all dimensional quantities in units of $M$, setting $M=1$.

For $\gamma = 3$, the metric reduces exactly to the Schwarzschild solution. When $\gamma < 4$, the geometry corresponds to a black hole with a single event horizon. As $\gamma$ approaches the critical value $\gamma = 4$, the configuration becomes extremal, with coincident horizons, and for $\gamma > 4$ the spacetime transitions into a traversable wormhole geometry.

The evolution of a test scalar field $\Phi$ in this background is governed by the covariant Klein–Gordon equation,
\begin{equation}\label{KGg}
\frac{1}{\sqrt{-g}}\,
\partial_{\mu}\!\left(\sqrt{-g}\, g^{\mu\nu}\partial_{\nu}\Phi\right)
= \mu^{2}\Phi,
\end{equation}
where $\mu$ is the scalar mass, $g_{\mu\nu}$ is the spacetime metric, and $g$ its determinant. This equation describes the propagation of massive scalar perturbations in curved spacetime.

Separating variables in the metric~(\ref{metric}) reduces the problem to a radial equation of Schrödinger type,
\begin{equation}\label{wave-equation}
\frac{d^{2}\Psi}{dr_{*}^{2}} + \left(\omega^{2} - V(r)\right)\Psi = 0,
\end{equation}
where $\Psi$ is the radial wave function, $\omega$ is the (generally complex) frequency, and $V(r)$ is the effective potential determined by the background geometry and the field properties. The tortoise coordinate $r_{*}$ is defined through
\begin{equation}\label{tortoise}
dr_{*} = \frac{B(r)}{f(r)}\,dr,
\end{equation}
which maps the event horizon to $r_{*} \to -\infty$ and facilitates the imposition of appropriate boundary conditions in quasinormal-mode analysis.

For scalar perturbations, the effective potential is given by
\begin{equation}\label{potentialScalar}
V(r) = f(r)\left(\mu^{2}+\frac{\ell(\ell+1)}{r^{2}}\right)
+ \frac{1}{r}\,\frac{d^{2}r}{dr_{*}^{2}},
\end{equation}
where $\ell=0,1,2,\ldots$ denotes the multipole number. Representative profiles of the effective potential for various values of $\mu$ and $\gamma$ are displayed in Fig.~\ref{fig:pot}. At large distances the potential asymptotically approaches $\mu^{2}$. Importantly, for the parameter range considered here the potential remains positive outside the event horizon, ensuring the absence of exponentially growing modes and thus the stability of scalar perturbations.

\section{Numerical and semi-analytical methods for finding quasinormal modes}

Here we will discuss two methods for calculation of QNMs: the approximate WKB method (used here only for checking of the results) and the Frobenius (Leaver) method allowing us to find quasinormal frequencies with any desired accuracy. 

\subsection{WKB approach}\label{sec:WKB}

If the effective potential $V(r)$ appearing in Eq.~(\ref{wave-equation}) exhibits a single-peak barrier structure, as shown in Fig.~\ref{fig:pot} for sufficiently small values of $\mu$, the Wentzel–Kramers–Brillouin (WKB) method provides a convenient and reliable technique for computing the dominant quasinormal frequencies. These frequencies are defined by imposing the standard quasinormal boundary conditions: purely ingoing waves at the event horizon and purely outgoing waves at spatial infinity.

The WKB construction relies on matching asymptotic solutions that satisfy the quasinormal boundary conditions with a local expansion of the effective potential in the vicinity of its maximum. At leading order, this procedure reduces to the well-known eikonal approximation, which becomes exact in the limit of large multipole number $\ell$. Away from the strict eikonal regime, the complex frequencies can be written as a series expansion around the peak of the potential \cite{Konoplya:2019hlu},
\begin{eqnarray}\label{WKBformula-spherical}
\omega^{2} &=& V_{0} + A_{2}(\K^{2}) + A_{4}(\K^{2}) + A_{6}(\K^{2}) + \ldots \\\nonumber
& & - i\K\sqrt{-2V_{2}}\left[1 + A_{3}(\K^{2}) + A_{5}(\K^{2}) + \ldots\right],
\end{eqnarray}
where $V_{0}$ and $V_{2}$ denote, respectively, the value of the potential and its second derivative with respect to the tortoise coordinate $r_{*}$ evaluated at the maximum. The coefficients $A_{i}(\K^{2})$ represent higher-order WKB corrections and depend on successive derivatives of the potential at that point.

The parameter $\K$ is related to the overtone number $n$ by
\begin{equation}
\K = n + \frac{1}{2}, \qquad n = 0,1,2,\ldots,
\end{equation}
so that the imaginary part of $\omega$ determines the decay rate of the corresponding QNM.

Closed-form expressions for the WKB correction terms are known up to high order. The second- and third-order formulas were obtained in \cite{Iyer:1986np}, while the extension to sixth order was developed in \cite{Konoplya:2003ii}. Subsequent work pushed the expansion up to thirteenth order \cite{Matyjasek:2017psv}. Since the WKB series is asymptotic rather than convergent, intermediate orders — typically the sixth or seventh — usually provide the optimal balance between accuracy and stability. The WKB formalism has been extensively employed in investigations of quasinormal spectra and grey-body factors across a wide range of black-hole geometries; see, for example, \cite{Konoplya:2001ji,Konoplya:2005sy,Tan:2022vfe,Momennia:2022tug,Skvortsova:2023zca,Skvortsova:2023zmj,Zhao:2022gxl,Bolokhov:2025fto,Kokkotas:2010zd,Konoplya:2006ar,Konoplya:2022hbl,Xiong:2021cth,Gong:2023ghh,Becar:2023zbl,Xia:2023zlf,Breton:2017hwe,Bolokhov:2025lnt,Dubinsky:2025fwv,Skvortsova:2024msa,Skvortsova:2024wly,Skvortsova:2024atk} for representative applications.

\subsection{Leaver method}

The second order differential equation can be cast into the form where all the coefficients have polynomial form.
\begin{eqnarray}
&&A(r) \Psi ''(r)+B(r) \Psi '(r)+C(r) \Psi (r)=0,
\\\nonumber&&
\\\nonumber
A(r)&=&r^2(3M-r)(r-2 M)^2 (\gamma  M-2 r),
\\\nonumber
B(r)&=&
M r (r-2 M) \Big(\gamma  \left(6 M^2-6 M
   r+r^2\right)
\\\nonumber&& +r (5 r-6 M)\Big),
\\\nonumber
C(r)&=&-\gamma M (r-2M)\left(6M^{2}-6Mr+r^{2}\right)
\\\nonumber&&
+r\Big(6M^{3}\left(3r^{2}\mu^{2}+3\ell(\ell+1)-2\right)
\\\nonumber&&
+M^{2}r\left(r^{2}\left(9\omega^{2}-33\mu^{2}\right)-33\ell(\ell+1)+16\right)
\\\nonumber&&
+Mr^{2}\left(4r^{2}\left(5\mu^{2}-3\omega^{2}\right)+5\left(4\ell(\ell+1)-1\right)\right)
\\\nonumber&&
+4r^{3}\left(r^{2}(\omega-\mu)(\omega+\mu)-\ell(\ell+1)\right)\Big).
\end{eqnarray}

To obtain highly accurate values of the quasinormal frequencies, we employ the Frobenius expansion technique in its continued-fraction formulation, commonly referred to as the Leaver method \cite{Leaver:1985ax}. This method is commonly recognized as one of the most precise semi-analytical tools for determining quasinormal spectra, especially effective in accurately capturing both the fundamental mode and higher overtones.

For a massive scalar field, the quasinormal boundary conditions read
\begin{equation}\label{bcond}
\Psi(r) \propto 
\begin{cases}
e^{-i \omega r_*}, & r_* \to r_h \quad (\text{event horizon}), \\
e^{\sqrt{\omega^2 - \mu^2 }\, r_*}, & r_* \to \infty,
\end{cases}
\end{equation}
where the solution is purely ingoing at the horizon and exponentially decaying at spatial infinity. Notice that $\omega$ always has non-zero real and imaginary parts,  so that the case of real $\omega^2 < \mu^2$  is excluded, while in the rest of the cases the sign of $\sqrt{\omega^2 -\mu^2}$  is chosen to stay in the same complex surface quadrant as $\omega$. 

These asymptotic behaviors are incorporated explicitly by extracting the dominant radial dependence and representing the remaining part of the solution as a generalized Frobenius series expanded about the horizon. Accordingly, we write
\begin{equation}
\Psi(r) = F(r)\sum_{n=0}^{\infty} a_n \left(\frac{r-r_h}{r}\right)^n ,
\end{equation}
with $r_h$ denoting the horizon radius. The prefactor $F(r)$ is constructed so as to satisfy the quasinormal boundary conditions and to ensure regularity of the series in the exterior region.

Substituting this expression into the radial equation produces a recurrence relation among the coefficients $a_n$. In many cases of physical interest, the resulting relation reduces to a three-term recurrence,
\begin{eqnarray}
\alpha_0 a_1 + \beta_0 a_0 &=& 0.\\\nonumber
\alpha_n a_{n+1} + \beta_n a_n + \gamma_n a_{n-1} &=& 0,
\qquad n \geq 1.
\end{eqnarray}
The coefficients $\alpha_n$, $\beta_n$, and $\gamma_n$ depend on the parameters of the black hole, the multipole number $\ell$, the scalar mass $\mu$, and the complex frequency $\omega$.

The requirement that the Frobenius series converge selects a discrete spectrum of admissible frequencies. This condition can be expressed via the equation with an infinite continued fraction,
\[
\frac{a_1}{a_0} = -\frac{\beta_0}{\alpha_0}
= \frac{\gamma_1}{
\beta_1 - \dfrac{\alpha_1 \gamma_2}{
\beta_2 - \dfrac{\alpha_2 \gamma_3}{
\beta_3 - \dfrac{\alpha_3 \gamma_4}{
\beta_4 - \cdots
}}}} ,
\]
whose numerical solution yields the quasinormal frequencies for given $\mu$, $\ell$, and background parameters.

In the present problem, the direct substitution of the Frobenius ansatz often leads to recurrence relations involving more than three neighboring coefficients. This situation typically arises when the radial equation contains additional structure or higher-order singular points. Such higher-order recurrences can be systematically reduced to an equivalent three-term form by Gaussian elimination \cite{Leaver:1986gd}, after which the standard continued-fraction procedure can be applied.

For highly damped modes or large overtone numbers, convergence of the continued fraction may deteriorate. Numerical stability can also be affected by irregular singularities in the complex plane. To mitigate these issues, we implement two standard improvements. First, we apply the method of \emph{integration through the midpoint} \cite{Rostworowski:2006bp}, which evaluates the series at an intermediate radial position and suppresses spurious divergences associated with asymptotic expansions. Second, we employ the \emph{Nollert improvement} \cite{Nollert:1993zz,Zhidenko:2006rs}, which analytically approximates the asymptotic behavior of the continued fraction and substantially accelerates convergence, particularly for higher overtones.

\begin{table}
\begin{tabular*}{\linewidth}{@{\extracolsep{\fill}}lcc}
\hline
$\mu$ & $\omega$ ($\gamma=0.46$) &  $\omega$ ($\gamma=0.4$)\\
\hline
\hline
 0 & 0.23007607-0.31001291 i & 0.23004562-0.31215392 i \\
 0.08 & 0.22920527-0.30789329 i &0.22915452-0.31005119 i\\
 0.16 & 0.22645007-0.30170057 i & 0.22634074-0.30391098 i \\
 0.24 & 0.22150744-0.29197840 i & 0.22130803-0.29428045 i \\
 0.32 & 0.21426069-0.27960451 i & 0.21394838-0.28203897 i \\
 0.40 & 0.20501544-0.26541830 i & 0.20457459-0.26802689 i \\
 0.48 & 0.19426818-0.24987606 i & 0.19369079-0.25270269 i \\
 0.56 & 0.18238968-0.23312707 i & 0.18167721-0.23621647 i \\
 0.64 & 0.16957124-0.21522071 i & 0.16873442-0.21861585 i \\
 0.72 & 0.15589250-0.19620563 i & 0.15495035-0.19994595 i \\
 0.80 & 0.14138111-0.17615173 i & 0.14035864-0.18027087 i\\
 0.88 & 0.12604554-0.15514274 i & 0.12496989-0.15967091 i  \\
 0.96 & 0.10988249-0.13326907 i & 0.10878663-0.13823621 i\\
 1.04 & 0.09288461-0.11063147 i & 0.09181028-0.11606165 i \\
 1.12 & 0.07505195-0.08733619 i & 0.07404433-0.09324343 i \\
 1.20 & 0.05639260-0.06348642 i & 0.05549592-0.06987687 i \\
 1.28 & 0.03691852-0.03918053 i & 0.03617648-0.04605493 i  \\
 1.33 & 0.02434288-0.02381331 i & 0.02371720-0.03097497 i \\
 1.37 & 0.01405062-0.01140961 i & 0.01354052-0.01882243 i\\
 1.40 & 0.00594754-0.00176286 i & 0.00578777-0.00966328 i\\
 1.401 & 0.00594754-0.00176286 i & 0.00552758-0.00935737 i \\
 1.402 & 0.00568414-0.00145095 i & 0.00526729-0.00905143 i \\
 1.403 & 0.00542062-0.00113901 i & 0.00500688-0.00874545 i\\
 1.404 & 0.00515698-0.00082703 i & 0.00474636-0.00843943 i\\
 1.405 & 0.00489322-0.00051502 i & 0.00448572-0.00813338 i\\
 1.406 & 0.00462935-0.00020298 i & 0.00422497-0.00782729 i \\
  1.406 & 0.00462935-0.00020298 i & 0.00422497-0.00782729 i \\
1.4061 & 0.00460295-0.00017177 i& 0.00419889-0.00779668 i \\
  1.4062 & 0.00457656-0.00014056 i& 0.00417281-0.00776607 i \\
  1.4063 & 0.00455016-0.00010935 i& 0.00414673-0.00773545 i \\
  1.4064 & 0.00452376-0.00007815 i& 0.00412064-0.00770484 i \\
  1.4065 & 0.00449736-0.00004694 i & 0.00409456-0.00767423 i \\ 
 1.410 & -- & 0.00318086-0.00660257 i \\
 1.411 & -- &0.00291954-0.00629630 i \\
 1.412 & -- &0.00265812-0.00598999 i \\
 1.413 & -- &0.00239659-0.00568365 i \\
 1.414 & -- &0.00213494-0.00537728 i \\
 1.415 & -- &0.00187318-0.00507087 i \\
 1.416 & -- &0.00161130-0.00476442 i \\
 1.417 & -- &0.00134932-0.00445794 i \\
 1.418 & -- &0.00108722-0.00415143 i \\
 1.419 & -- &0.00082501-0.00384488 i \\
 1.420 & -- &0.00056269-0.00353829 i \\
 1.421 & -- &0.00030026-0.00323168 i \\
 1.422 & -- &0.00003771-0.00292502 i \\
 1.4221 & --&0.00001145-0.00289436 i \\
 \hline
 \hline
\end{tabular*}
\caption{Fundamental QNMs $\ell=n=0$ for various values of $\gamma$ near the threshold at which the vanishing real part is changed by the vanishing damping rate.}\label{tabl:data}
\end{table}

With these refinements, the Leaver method provides a highly precise and robust framework for computing quasinormal spectra. It has been successfully applied to both massless and massive perturbations in a wide range of black-hole geometries \cite{Rosa:2011my,Konoplya:2007zx,Dubinsky:2024fvi,Konoplya:2004uk,Konoplya:2023ahd,Saka:2025xxl}. Owing to its numerical accuracy, this approach is especially well suited for resolving subtle spectral features such as quasi-resonances and long-lived modes characteristic of massive fields.

\section{Quasinormal Modes}

The QNMs were computed using the convergent Leaver method, which provides highly accurate numerical results. The WKB approximation was employed to obtain an initial estimate of the frequencies in the massless limit, serving as a starting guess for the subsequent Leaver iteration.

The quasinormal spectrum of a massive scalar field in the CFM brane-world black-hole background exhibits two qualitatively distinct behaviors depending on the field mass $\mu$ and the tidal parameter $\gamma$.

\begin{figure*}
\resizebox{\linewidth}{!}{\includegraphics{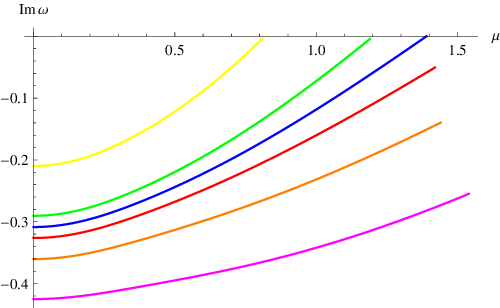}~~\includegraphics{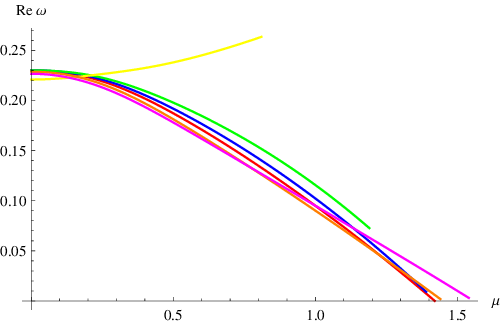}}
\caption{Real (left) and imaginary (right) parts of the fundamental QNM $\ell=n=0$ for  $\gamma=3$ (yellow), $\gamma=1$ (green), $\gamma=0$ (red), $\gamma=0.5$ (blue), $\gamma=-1$ (orange), $\gamma=-3$ (magenta).}\label{fig:L0n0}
\end{figure*}

\begin{figure*}
\resizebox{\linewidth}{!}{\includegraphics{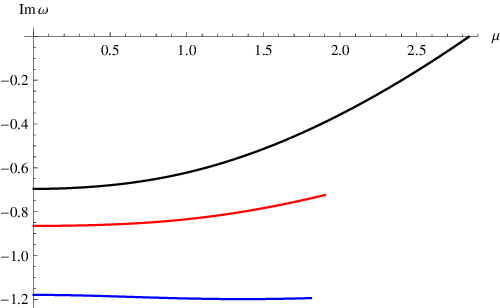}~~\includegraphics{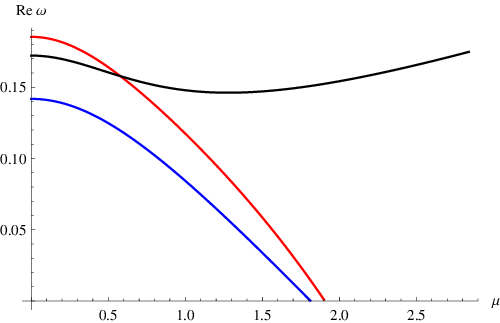}}
\caption{Real (left) and imaginary (right) parts of the first overtone $\ell=0$, $n=1$ for  $\gamma=3$ (black), $\gamma=2$ (red), $\gamma=0$ (blue), $\gamma=0.5$ (blue).}\label{fig:L0n1}
\end{figure*}

Figure~\ref{fig:L0n0} displays the dependence of the fundamental mode ($\ell=n=0$) on the scalar mass for several values of $\gamma$. Two regimes are clearly observed.

First, for sufficiently small positive and for all negative values of $\gamma$, the real part of the frequency decreases as $\mu$ increases and eventually approaches zero. In this case the oscillatory component of the mode disappears, leaving a purely imaginary frequency. This transition point depends sensitively on the value of $\gamma$, as illustrated in Table~\ref{tabl:data}, where the fundamental frequency is tracked near the threshold at which the real part vanishes.

Second, for larger values of $\gamma$, a different behavior is found: as the scalar mass increases, the imaginary part of the frequency (the damping rate) tends toward zero instead. In this regime the modes become increasingly long-lived. The data in Table~\ref{tabl:data} show how, near the transition between these two behaviors, the character of the fundamental mode changes: for close values of $\gamma$, either the real part or the imaginary part approaches zero first, signaling a switch between the two types of spectral evolution.

An important feature emphasized in the manuscript is that when either the real or the imaginary part of a given mode reaches zero, that particular mode ceases to exist in the spectrum and is replaced by the first overtone. Thus, the spectrum reorganizes itself as $\mu$ varies.

Overall, the analysis demonstrates that introducing a mass term for the scalar field qualitatively modifies the quasinormal spectrum of the CFM black hole. Depending on the value of the tidal parameter $\gamma$, the system supports either modes whose oscillation frequency vanishes or modes whose damping rate vanishes in the large-$\mu$ regime, leading to two distinct spectral branches.

As the tidal parameter $\gamma$ varies, the spectrum undergoes a transition from modes whose real part vanishes to modes whose imaginary part tends to zero. The numerical data presented in Table~\ref{tabl:data} suggest the existence of a threshold value of $\gamma$ at which both parts approach zero simultaneously; however, establishing this behavior with high numerical accuracy proves to be technically challenging.

We also observe that the qualitative behavior of the spectrum at higher multipole numbers $\ell$ and higher overtones remains similar to that described for the fundamental case. However, the transition to regimes where either the real part or the imaginary part of the frequency vanishes occurs at larger values of the scalar-field mass $\mu$ as $\ell$ or $n$ increases (see Fig. \ref{fig:L0n1}). Thus, higher multipoles require a stronger mass contribution to exhibit the same type of spectral reorganization. 

In addition, the threshold value of the tidal parameter $\gamma$ at which the behavior changes from modes with vanishing real part to modes with vanishing imaginary part shows a pronounced dependence on the multipole number. This indicates that the competition between the two spectral branches is sensitive not only to the field mass but also to the angular structure of the perturbation.

Here we have presented results primarily for the fundamental mode and the first overtone in order to demonstrate that, when the fundamental mode approaches the quasi-resonant regime and effectively disappears from the spectrum, the first overtone becomes the dominant (least damped) mode. However, the convergent Frobenius (Leaver) method allows one to compute QNMs with arbitrarily high accuracy, limited only by computational resources. One can therefore extend the analysis further and show that, as higher overtones successively acquire vanishing real or imaginary parts, the next overtone in the sequence takes over as the least damped (dominant) mode, and so on. The overtones are highly sensitive to small static deformations of the near-horizon geometry \cite{Konoplya:2022pbc}, so that, in principle, the overtone spectrum of the CFM background may differ significantly from its Schwarzschild counterpart even for massless fields. Nevertheless, the excitation factors of overtones are typically suppressed compared to the fundamental mode, rendering them practically invisible in the time-domain profiles.

\section{Conclusions}\label{sec:conclusions}

In this work, we have investigated the QNMs of a massive scalar field in the background of the CFM brane-world black hole. Unlike the previous study of the problem  \cite{Abdalla:2006qj,Bronnikov:2019sbx,Malik:2024itg} where time-domain integration and WKB methods were used, in the present work the spectrum was computed using the convergent Leaver method, while the WKB approximation was employed to obtain initial frequency estimates in the massless limit. This combination allowed us to reliably trace the evolution of the modes as the scalar mass $\mu$ and the tidal parameter $\gamma$ vary and find qualitatively new behavior.

We have shown that the quasinormal spectrum exhibits two qualitatively distinct types of behavior. For certain values of the tidal parameter, the real part of the frequency decreases with increasing $\mu$ and eventually vanishes, leading to non-oscillatory modes. In another regime, the imaginary part of the frequency tends to zero, producing increasingly long-lived modes. When either the real or imaginary part reaches zero, the corresponding mode disappears from the spectrum and is replaced by the first overtone, indicating a reorganization of the spectral structure as the field mass increases.

These results demonstrate that the presence of a scalar-field mass significantly modifies the quasinormal spectrum of the CFM black hole and that the tidal parameter $\gamma$ plays a crucial role in determining the qualitative behavior of the modes. The interplay between the field mass and the brane-world deformation parameter leads to distinct spectral branches, which may provide further insight into the dynamical properties of such geometries.

\section*{Data Availability Statement}
All data generated or analyzed during this study are included in this published article.

\section*{Conflict of Interest}
The authors declare no conflict of interest.

\vspace{5mm}
\begin{acknowledgments}
B. C. L. is grateful to the Excellence project FoS UHK 2205/2025-2026 for the financial support.
\end{acknowledgments}

\bibliography{bibliographyXX}
\end{document}